\documentstyle[12pt]{article}

\begin{document}
\begin{titlepage}
\begin{flushright}
MRI-PHY/96-16 \\ 
hep-th/9606009
\end{flushright}
\vskip 1.4cm
\begin{center}

{\Large \bf F-theory from Dirichlet 3-branes}

\vspace{4ex}

{\large Dileep P. Jatkar and S. Kalyana Rama}


Mehta Research Institute of Mathematics \& Mathematical Physics\\
10, Kasturba Gandhi Marg, Allahabad 211 002, India. 

email: dileep, krama@mri.ernet.in \\ 

\end{center}
\vspace{4ex}
\abstract{
Starting from the type IIB Dirichlet 3-brane action, we obtain 
a Nambu-Goto action. It is interpreted as the world volume 
action of a fundamental 3-brane, and its target space theory 
as F-theory. The target space is twelve dimensional, with 
signature $(11, 1)$. It is an elliptic fibration over a ten 
dimensional base space. The $SL(2, Z)$ symmetry of type IIB 
string has now an explicit geometric interpretation. Also, 
one gets a glimpse of the conjectured self-duality 
of M-theory. 
}
\vfill
\end{titlepage}

\section{Introduction}

The strong-weak coupling duality in the string theory 
\cite{font}-\cite{sen} has shed a lot of light on 
the non-perturbative physics of string theory. In particular, 
different string theories are related to each other by these 
duality transformations \cite{hultow}-\cite{schwarz}. 
It was also realised 
that the RR sector of type II strings contains 
a wealth of information about the nonperturbative aspects of 
type II strings. However, the solutions carrying RR charges do 
not appear in the elementary excitation spectrum and, hence, 
there was no satisfactory understanding of the RR charged 
objects. But, recently, in a beautiful paper, Polchinski 
\cite{pol} has shown that these solitons with RR charges are 
precisely the Dirichlet branes, much studied earlier 
on \cite{polnotes}. From 
the burst of activity that followed this discovery, one now 
has a deeper understanding of the duality symmetries of 
various strings, and the relations among them 
\cite{wit}-\cite{bsv}. 

One ramification of these developments is the emergence of 
an unifying eleven dimensional theory, referred to as 
M-theory, which describes the target space theory 
of fundamental 
2-branes. All known string theories are obtained from M-theory 
after appropriate dimensional reductions. The geometric origin 
of many of the duality symmetries is also clear in M-theory 
\cite{witten,schwarz,horwit}. One exception is type IIB 
strings, whose connection to M-theory and to other strings is 
understood only after compactifying these theories to nine, or 
lower, dimensions \cite{schwarz}. Also unknown is 
the geometric origin of type IIB duality symmetries. 

Now, from the works of \cite{h}-\cite{bds}, there appears to 
exist a twelve dimensional theory, dubbed F-theory, which is 
likely to describe the target space theory 
of fundamental 3-branes. F-theory 
is only now beginning to be explored, but it already holds 
many promises. It appears to incorporate type IIB strings 
naturally and provides new insights 
upon compactification to lower dimensions. Quite excitingly, 
its compactification to four dimensions can provide 
a realisation, as pointed out in \cite{v}, of Witten's novel 
proposal towards solving the cosmological constant 
problem \cite{wittencc}. 

Returning to M-theory, the appearance of the eleven 
dimensional spacetime can also be seen from a different point 
of view. Type II strings have Dirichlet $p$-branes, with $p$ 
even (odd) for type IIA (B) \cite{polnotes,duffpr,towni}.  
Starting from the Dirichlet 2-brane action 
\cite{polnotes,callan,s}, a Nambu-Goto action for 
a fundamental 
2-brane has been obtained in \cite{s,t} (also see \cite{to}). 
The target space of the fundamental 2-brane turns out to be 
eleven dimensional, which is the M-theory spacetime. For 
$p = 1$, this method gives the ten dimensional target space of 
the fundamental string \cite{s}. 

Therefore, one may naturally hope that starting from 
the Dirichlet 3-brane action, a Nambu-Goto action can be 
obtained which can be interpreted as the world volume action 
of a fundamental 3-brane, and its target space theory as 
F-theory. The F-theory thus derived is likely to offer 
insights into the duality symmetries of type IIB strings. 
3-branes are also special for another reason. 
Any $p$-brane of type II strings can be mapped, by Poincare 
duality, to a $(6 - p)$-brane. Thus, $(p > 3)$-branes can be 
equivalently described by $(p < 3)$-branes. However, 3-branes 
are mapped onto 3-branes themselves and, hence, they must be 
considered on their own right. 

Dirichlet 3-branes have, in fact, been considered in \cite{t}. 
However, it was found that applying the techniques of 
\cite{s,t} to the Dirichlet 3-brane action gives back 
the original action only, in terms of a dual gauge field. It 
is not a Nambu-Goto action, and has no interpretation as 
the theory of a fundamental 3-brane and, hence, 
does not lead to F-theory. 

In this paper, we study the 3-branes. We start from the type 
IIB Dirichlet 3-brane action and adopt a simple but 
direct approach, in some ways analogous to that of 
\cite{s,t}. Namely, we first perform a double dimensional 
reduction and then apply the methods of \cite{s,t} to 
the resulting action. We find the following: 

The action thus obtained is a Nambu-Goto action, and can be 
interpreted as the world volume action of a fundamental 
3-brane. The target space is  
twelve dimensional, with signature $(11, 1)$. It is locally 
a product of a ten dimensional spacetime and a torus, with 
a fixed K\" ahler class. That is, the twelve dimensional target 
space is an elliptic fibration, having the structure of 
a fiber bundle whose fiber is a two dimensional torus over 
a ten dimensional base space \cite{v}. 

The $SL(2, Z)$ symmetry of type IIB string has an explicit 
geometric interpretation: the $SL(2, Z)$ 
modulus of type IIB string is the complex structure 
modulus of the torus fiber.\footnote{That such a geometric 
interpretation of the $SL(2, Z)$ symmetry of type IIB theory 
exists was anticipated in \cite{schwarz,v}.} 
The strong-weak coupling duality of type IIB exchanges 
the coordinates of the torus. Also, the two radii of 
the torus are proportional to $e^{-\phi}$ and $e^{\phi}$, 
where $e^{- \phi}$ is the string coupling. Thus, 
both in the strong and weak coupling limit, 
the twelve dimensional theory appears to be eleven 
dimensional with (10,1) signature. Perhaps, this is related 
to the conjectured self-duality of M-theory \cite{dvv}. 

Because of the double dimensional reduction, the Nambu-Goto 
action for the 3-brane appears in a gauge fixed form, where 
two conditions need to be imposed. However, this is a generic 
phenomenon, as will be shown by a simple counting argument. 
That is, two such conditions are always necessary if a 3-brane 
Nambu-Goto action is derivable from the Born-Infeld action of 
a Dirichlet 3-brane. The resulting target space action does 
not have the full twelve dimensional general coordinate 
invariance. Such a possibility has been mentioned in \cite{t}, 
as perhaps a way of circumventing Nahm's theorem regarding 
realisations of supersymmetry in spacetime with dimensions 
$> 11$ \cite{nahm}. 

The paper is organised as follows. In section 2, we start 
from the Born-Infeld action of a type IIB Dirichlet 3-brane, 
and derive the Nambu-Goto action for a fundamental 3-brane. 
In section 3, we obtain the target space metric of 
the fundamental 3-brane, in a form where the duality 
symmetries become explicit. In section 4, we describe 
the origin of the duality symmetry of IIB strings. In 
section 5, we conclude with a brief discussion. 

\section{Nambu-Goto Action }

In the following, we study the Dirichlet (D-) 3-brane of 
the ten dimensional type IIB string. In our notation, 
$X^\mu$ denotes the coordinates 
of the ten dimensional target space of type IIB string and 
$\xi^\alpha$, those of the 3-brane world volume. The massless 
fields of the type IIB string are the following: a dilaton 
$\phi$, a metric tensor $g_{\mu \nu}$, and a second rank 
antisymmetric tensor $B_{\mu \nu}$ in Neveu-Schwarz (NS) 
sector; $\chi$, the antisymmetric tensors $C_{\mu \nu}$, and 
$A_{\mu \nu \rho \sigma}$ of rank $2$ and $4$ respectively in 
RR sector. These target space tensor fields induce 
$g_{\alpha \beta}, \; B_{\alpha \beta}, \; C_{\alpha \beta}$, 
and $A_{\alpha \beta \gamma \delta}$ on the D-3-brane. 
For example, 
\begin{equation}\label{indmet}
g_{\alpha \beta} = \partial_{\alpha} X^{\mu} \partial_{\beta} 
X^{\nu} g_{\mu \nu} \; . 
\end{equation}
The RR charges are carried 
by D-instanton, D-string and D-3-brane. In the presence of 
the D-branes, there will also be a boundary gauge field 
$A_{\alpha}$ in the NS sector. Also, define 
\begin{equation}\label{fk}
{\cal F}_{\alpha \beta} = 
\partial_{\alpha} A_{\beta} - \partial_{\beta} A_{\alpha} 
+ \frac{1}{2} (B_{\alpha \beta} - B_{\beta \alpha}) \; . 
\end{equation} 

The action for the D-3-brane of the type IIB string 
can be written as \cite{s,t} 
\begin{equation}\label{bi}
S_{BI} = \int d^4 \xi \left( e^{- \phi} \sqrt{-g P} + Q 
\right)  \; , 
\end{equation}
where $g = {\rm det} (g_{\alpha \beta}) , \; 
\alpha, \beta, \ldots \in \{0, 1, 2, 3\}$, 
\begin{eqnarray}
P & = & {\rm det} ({\bf 1} +  {\cal F}_\alpha ^\beta) 
= 1 + \frac{1}{2} {\cal F}_{\alpha \beta} 
{\cal F}^{\alpha \beta} - \frac{1}{64} 
( \epsilon^{\alpha \beta \gamma \delta} 
{\cal F}_{\alpha \beta} {\cal F}_{\gamma \delta} )^2 
\nonumber \\ 
Q & = & \frac{\epsilon^{\alpha \beta \gamma \delta}}{24} 
( A_{\alpha \beta \gamma \delta} + 6 C_{\alpha \beta} 
{\cal F}_{\gamma \delta} + 3 \chi {\cal F}_{\alpha \beta} 
{\cal F}_{\gamma \delta} ) \; ,  \label{pq} 
\end{eqnarray}
and $g^{\alpha \beta}$ is used to raise the indices. 

Note that the term $e^{- \phi} \sqrt{-g P}$ in (\ref{bi}) 
can be written as 
\begin{equation}\label{phis}
\sqrt{- {\rm det} (e^{- \frac{\phi}{2}} g_{\alpha \beta} 
+  e^{- \frac{\phi}{2}} {\cal F}_{\alpha \beta} ) } \; .
\end{equation}
Now, if one makes the replacements  
\begin{eqnarray}
g_{\alpha \beta}& \to & e^{- \frac{\phi}{2}} g_{\alpha \beta} 
\nonumber \\ 
{\cal F}_{\alpha \beta} & \to & e^{- \frac{\phi}{2}} 
{\cal F}_{\alpha \beta}  \nonumber \\ 
C_{\alpha \beta} & \to & e^{\frac{\phi}{2}} 
C_{\alpha \beta}  \nonumber \\ 
\chi & \to & e^{\phi} \chi \; , \label{replace} 
\end{eqnarray}
then the resulting action, in terms of the replaced variables, 
is the same as that in (\ref{bi}) but with $\phi = 0$. 
Therefore, we set $\phi = 0$ in this section, and restore 
its dependence in the next section. 

At this stage, following \cite{s,t}, one may introduce 
a Lagrange multiplier enforcing equation (\ref{fk}) as 
a constraint and, then, eliminate both $A_{\alpha}$ and 
${\cal F}_{\alpha \beta}$, now treated as independent fields. 
For D-2-brane, the resulting action is the Nambu-Goto action of 
a fundamental 2-brane, with eleven dimensional target space 
\cite{s,t}. The dual of $A_{\alpha}$ is the extra world volume 
scalar, which becomes the eleventh target space coordinate. 

For 3-brane, this procedure does not lead to the Nambu-Goto 
action, but leads back to the action (\ref{bi}), now in 
terms of a dual gauge field \cite{t}.
\footnote{Actually, this is so only after the addition of 
a term $\propto \epsilon^{\alpha \beta \gamma \delta} 
B_{\alpha \beta} C_{\gamma \delta}$ which, however, spoils 
the invariance of (\ref{bi}) under $\delta B_{\alpha \beta} = 
\partial_\alpha \Lambda_\beta - \partial_\beta \Lambda_\alpha, 
\; \; \delta A_\alpha = - \Lambda_\alpha$. To restore this 
invariance, one needs to postulate an anamolous transformation 
law for $A_{\alpha \beta \gamma \delta}$, of the type 
described in \cite{sch}. } But, this action 
is not in the Nambu-Goto form and, hence, cannot be 
interpreted as a world volume action of a fundamental 3-brane. 

Therefore, we proceed differently at this stage. We first 
perform a double dimensional reduction, by setting $\xi^3=X^9$ 
and taking the fields to be independent of this coordinate. 
We also define, for the sake of convenience, 
the following: 
\begin{eqnarray}
& & A_3 = U \; , \; \;  
b_m = \partial_m U + B_m\; ,\; 
A_{l m n} = A_{l m n 3} \; , \nonumber \\ 
& & B_m = \frac{1}{2} (B_{m 3} - B_{3 m}) \; , \; \; 
C_m = \frac{1}{2} (C_{m 3} - C_{3 m}) \; ,  \label{vbc}
\end{eqnarray}
where $l, m, \ldots \in \{ 0, 1, 2 \}$. 
Then, after a straight forward algebra, various terms 
in (\ref{pq}) become: 
\begin{eqnarray}
{\cal F}_{\alpha \beta} {\cal F}^{\alpha \beta}  
& = & {\cal F}_{m n} {\cal F}^{m n} + 2 b_m b^m \nonumber \\ 
\epsilon^{\alpha \beta \gamma \delta} 
{\cal F}_{\alpha \beta} {\cal F}_{\gamma \delta} 
& = & 4 \epsilon^{l m n} b_l {\cal F}_{m n} \nonumber \\
\epsilon^{\alpha \beta \gamma \delta} 
A_{\alpha \beta \gamma \delta} 
& = & 4 \epsilon^{l m n} A_{l m n} \nonumber \\
\epsilon^{\alpha \beta \gamma \delta} 
C_{\alpha \beta} {\cal F}_{\gamma \delta}
& = & 2\epsilon^{l m n} (C_l {\cal F}_{m n} + b_l C_{m n}) 
\; . \label{012} 
\end{eqnarray} 
where, now, $g^{m n}$ is used to raise the indices and 
\begin{equation}\label{f}
{\cal F}_{m n} = \partial_m A_n - \partial_n A_m 
+ \frac{1}{2} (B_{m n} - B_{n m}) \; . 
\end{equation}
Then, $P$ and $Q$ in (\ref{pq}) become 
\begin{eqnarray}
P & = & 1 + b_m b^m + \frac{1}{2} {\cal F}_{m n} 
{\cal F}^{m n}  - \frac{1}{4} ( \epsilon^{l m n} b_l 
{\cal F}_{m n} )^2 \nonumber \\ 
Q & = & \frac{\epsilon^{l m n}}{6} \left( A_{l m n} 
+ 3 (C_l + \chi b_l) {\cal F}_{m n} + 3 b_l C_{m n} \right) 
\label{p1q1} \; . 
\end{eqnarray} 

Now, as in \cite{s,t}, we introduce a Lagrange multiplier 
$\Lambda^{m n} (= - \Lambda^{n m})$ enforcing equation 
(\ref{f}) as a constraint and, then, eliminate both $A_m$ 
and ${\cal F}_{m n}$, now treated as independent fields. 
The total action, including the Lagrange multiplier term, is 
given by 
\begin{equation}\label{slm}
S = L \int d^3 \xi \left( \sqrt{- g P} + Q 
- \frac{\Lambda^{m n}}{2} ({\cal F}_{m n} - \partial_m A_n 
+ \partial_n A_m 
- \frac{1}{2} (B_{m n} + B_{n m}) ) \right) \; . 
\end{equation}
where $P$ and $Q$ are given by (\ref{p1q1}), and 
$L = \int d \xi^3$ is the length of the $\xi^3$ coordinate. 
(Note that the integrand is independent of $\xi^3$ and, hence, 
$d \xi^3$ integral is trivial to perform.) Note that when 
restoring $\phi$-dependence, one should also make the replacement 
\begin{equation}\label{replace2}
\Lambda^{m n} \to e^{\frac{\phi}{2}} \Lambda^{m n} \; , 
\end{equation} 
along with the others given in (\ref{replace}). Eliminating 
$A_m$ from (\ref{slm}) now gives $\partial_m \Lambda^{m n} 
= 0$, which is solved identically by 
\begin{equation}\label{Lambda}
\Lambda^{m n} = \epsilon^{l m n} \partial_l V \; , 
\end{equation}
where $V$ is a scalar, dual to $A_m$. 

Collecting the terms linear in ${\cal F}_{m n}$ and defining 
$\lambda_m = \partial_m V - C_m - \chi b_m$, and 
\[
\hat{A}_{l m n} = A_{l m n} + 3 b_{[l} C_{m n]} 
+ 6 \partial_{[l} V B_{m n]} 
\] 
where $[..]$ denotes total antisymmetrisation with respect 
to the enclosed indices, we can write the action 
(\ref{slm}) as 
\begin{equation}\label{int}
S = L \int d^3 \xi \left( e^{- \phi} \sqrt{- g P} 
- \frac{\epsilon^{l m n}}{2} \lambda_l {\cal F}_{m n} 
+ \frac{\epsilon^{l m n}}{6} \hat A_{l m n} \right) \; . 
\end{equation} 
Now, it is straightforward to eliminate the field 
${\cal F}_{m n}$ also. Varying the action (\ref{int}) 
with respect to ${\cal F}_{m n}$ gives 
\[
\sqrt{- g} ({\cal F}^{m n} - \frac{1}{2} \epsilon^{l m n} 
\epsilon^{p q r} b _l b_p {\cal F}_{q r}) = 
- 2 \epsilon^{l m n} \lambda_l \sqrt{P} \; . 
\]
Multiplying this equation once by ${\cal F}_{m n}$, 
once by $\epsilon_{k m n} b^k$, and squaring each side of 
the above expression, result in three different equations. 
After some manipulations of these equations, ${\cal F}_{m n}$ 
can be completely eliminated from the action in (\ref{int}). 
The action (\ref{int}) can, finally, be written as 
\begin{equation}\label{int2}
S = L \int d^3 \xi \left( \sqrt{- g {\cal P}} 
+ \frac{\epsilon^{lmn}}{6}\hat A_{lmn} \right) \; , 
\end{equation}
where 
\[
{\cal P} = 1 + b_m b^m + \lambda_m \lambda^m 
+ (b_m b^m) (\lambda_n \lambda^n) - (b_m \lambda^m)^2 
\]
is nothing but a determinant ! That is, 
\[
{\cal P} = {\rm det}({\bf 1} + b_m b^n 
+ \lambda_m \lambda^n) \; , 
\]
which can be seen simply by evaluating the determinant. 
The above expression can equivalently be written as 
\begin{equation}\label{calp}
{\cal P} = {\rm det}({\bf 1} + b_\alpha b^\beta 
+ \lambda_\alpha \lambda^\beta) \; , 
\end{equation}
with $\alpha, \beta = 0, 1, 2, 3$, and $b_3 = \lambda_3 = 0$. 

Hence, the action (\ref{int}), or equivalently (\ref{int2}), 
which is the dimensionally reduced type IIB Dirichlet 3-brane 
action, can be written as 
\begin{equation}\label{ngterm}
S = \int d^4 \xi \left( \sqrt{- {\rm det}(g_{\alpha \beta} 
+ b_\alpha b_\beta + \lambda_\alpha \lambda_\beta) } 
+ \frac{\epsilon^{\alpha \beta \gamma \delta}}{6}
\hat A_{\alpha \beta \gamma \delta} \right) \; , 
\end{equation}
where 
\begin{eqnarray}
b_\alpha & = & \partial_\alpha U + B_\alpha \; , \nonumber \\
\lambda_\alpha & = & \partial_\alpha V - C_\alpha 
- \chi b_\alpha \; , \nonumber \\
\hat{A}_{\alpha \beta \gamma 3} 
& = & A_{\alpha \beta \gamma 3} 
+ 3 b_{[\alpha} C_{\beta \gamma]}
+ 6 \partial_{[\alpha} V B_{\beta \gamma]} \; , \label{b012}
\end{eqnarray}
for $\alpha, \beta, \ldots = 0, 1, 2$ and 
\begin{equation}\label{b3}
b_3 = \lambda_3 = 0 \; . 
\end{equation}
In this form, the action (\ref{ngterm}) can be interpreted 
as the Nambu-Goto action for a fundamental 3-brane. 
The term involving $\hat A$ is the Wess-Zumino term. 

It is clear from the explicit expressions of
$b_\alpha$ and $\lambda_\alpha$, that the target 
space of the fundamental 3-brane is twelve dimensional, with 
$U$ and $V$ giving rise to two extra target space 
coordinates. It is evident that both of these coordinates 
are spacelike and, thus, the signature of the target space 
that emerges here naturally is $(11, 1)$(see section 3). 
Moreover, the target space theory obtained from the Nambu-Goto 
action in (\ref{ngterm}) can be interpreted as the F-theory. 

Note that because of the double dimensional reduction 
procedure we adopted here, 
the Nambu-Goto action for the 3-brane in (\ref{ngterm}) 
appears in a gauge given by (\ref{b3}). Hence, it does 
not have the full twelve dimensional general coordinate 
invariance. However, this is likely to be a generic 
phenomenon. That is, two such conditions are always 
necessary, if a Nambu-Goto action were to be derivable from 
the Born-Infeld action of type IIB Dirichlet 3-brane. 
This can be seen by a simple counting as follows. 

The D-brane action of the type II strings 
already has the metric $g_{\alpha \beta}$ induced 
by ten target space coordinates, see (\ref{indmet}). Now, if 
the target space has two extra coordinates, as expected for 
a fundamental 3-brane, then there should be 8 (= number of 
extra target space coordinates $\times$ the world volume 
dimension) extra degrees of freedom, corresponding to 
$\partial_\alpha X^{10}$ and $\partial_\alpha X^{11}, \; 
\alpha = 0, 1, 2, 3$. The type IIB D-3-brane does have extra 
degrees of freedom on the world volume, coming from 
${\cal F}_{\alpha \beta}$, but they are 6 only in number, 
falling short by 2! This is the origin of 
the two conditions in (\ref{b3}). 

In the case of 2-branes, the target space is eleven 
dimensional, and the world volume is three dimensional, 
so the required number of extra degrees of freedom is 3. 
The type IIA D-2-brane has extra degrees of freedom on 
the world volume, coming from ${\cal F}_{m n}$, which are 
also 3 in number, exactly what is needed. Then, our counting 
argument suggests that a Nambu-Goto action for 
a fundamental 2-brane should be derivable from the type IIA 
Dirichlet 2-brane action, without any gauge condition. This 
is indeed the case, as shown in \cite{s,t}. 

Perhaps, instead of the type IIB Dirichlet 3-brane action, 
if a more general or an altogether different action with 
more degrees of freedom is used, then may be a Nambu-Goto 
action for a fundamental 3-brane is derivable without any 
gauge condition. However, we will not pursue it here 
(see discussion in section 5). 

That such a phenomenon generically occurs, and that 
the resulting target space action does not have the full 
twelve dimensional general coordinate invariance, may 
actually be a boon in disguise. As mentioned in \cite{t}, 
this may perhaps be related to the way F-theory implements 
supersymmetry, circumventing Nahm's theorem regarding 
realisations of supersymmetry in spacetime with dimensions 
$> 11$ \cite{nahm}. 

\section{Target space metric}

The twelve dimensional target space metric of the F-theory 
is easy to read off from (\ref{ngterm}). Before doing so, let 
us now restore the $\phi$-dependence. We only need to use 
equations (\ref{replace}) and (\ref{replace2}). Then, as can 
be seen easily, $b_\alpha$ and $\lambda_\alpha$ given in 
(\ref{b012}) are to be replaced as follows: 
\begin{equation}
b_\alpha \to e^{- \frac{\phi}{2}} b_\alpha \; , \; \; \; 
\lambda_\alpha \to e^{\frac{\phi}{2}} \lambda_\alpha \; . 
\end{equation}
The action (\ref{ngterm}), after these replacements, 
can be written as  
\begin{equation}\label{ngterm2}
S = \int d^4 x \left( 
\sqrt{- {\rm det} (\hat{g}_{\alpha \beta})} 
+ \frac{\epsilon^{\alpha \beta \gamma \delta}}{6}
\hat A_{\alpha \beta \gamma \delta} \right) \; , 
\end{equation}
where 
\begin{equation}\label{ghat}
\hat{g}_{\alpha \beta} = 
e^{- \frac{\phi}{2}} g_{\alpha \beta} 
+ e^{- \phi} b_\alpha b_\beta 
+ e^{\phi} \lambda_\alpha \lambda_\beta \; . 
\end{equation}
The action in (\ref{ngterm2}) now has the correct 
$\phi$-dependence. 

We now rewrite $\hat{g}_{\alpha \beta}$, in a form which 
makes explicit the origin of type IIB duality symmetries. 
Let us define a two component vector ${\bf Z_\alpha}$ and 
a matrix ${\cal M}$ as follows: 
\begin{equation}\label{curlym}
{\bf Z_\alpha} = 
\pmatrix{ \partial_\alpha V - C_\alpha \cr 
          \partial_\alpha U + B_\alpha     } \; , \; \;\; 
{\cal M} = \frac{1}{\lambda_2} 
\pmatrix{ 1         & \lambda_1          \cr 
          \lambda_1 & |\lambda|^2    } \; , 
\end{equation}
where $\lambda = \lambda_1 + i \lambda_2 \equiv 
\chi + i e^{- \phi} $. Then 
\begin{equation}\label{strmet}
\hat{g}_{\alpha \beta} = 
e^{- \frac{\phi}{2}} g_{\alpha \beta} 
+ {\bf Z_\alpha}^T {\cal M} {\bf Z_\beta} \; . 
\end{equation}

Similarly, let us now define 
\begin{eqnarray}
d U^1 & = & d V - \frac{1}{2} (C_{\mu 9} - C_{9 \mu}) 
d X^\mu \; , \nonumber \\ 
d U^2 & = & d U + \frac{1}{2} (B_{\mu 9} - B_{9 \mu}) 
d X^\mu \; . \label{u1u2}
\end{eqnarray}
Note that these equations for $U^1$ and $U^2$ can be solved 
locally and, depending on the topological properties of 
$B_{\mu \nu}$ and $C_{\mu \nu}$, may also be solved globally. 
The twelve dimensional target space line element 
is then given by 
\begin{equation}\label{ds12}
d s^2 = e^{-\phi/2}g_{\mu \nu} d X^\mu d X^\nu 
+ e^{\phi}  (d U^1)^2 + 2 e^{\phi} \chi dU^1 d U^2 
+ (e^{- \phi} + e^{\phi} \chi^2) (d U^2)^2 \; . 
\end{equation}
In this form, it is clear that $U^1$ and $U^2$ are the two 
extra target space coordinates, and that both of them are 
spacelike. Furthermore, as can be seen 
easily, the metric on this internal two dimensional space is 
given by 
\begin{equation}\label{gint}
g_{\rm int} = {\cal M} \; , 
\end{equation} 
where ${\cal M}$ is defined in (\ref{curlym}). This metric 
describes a torus, whose two radii of are 
proportional to $e^{-\phi}$ and $e^{\phi}$. Thus, locally, 
the twelve dimensional target space of F-theory is a product 
of a ten dimensional spacetime and a torus, with a fixed 
K\" ahler class (see also section 4). That is, the twelve 
dimensional space is an elliptic fibration over a ten 
dimensional base \cite{v}. The duality properties of this 
torus and its relation to the type IIB string 
duality symmetries will be discussed now. 

\section{Geometric Interpretation of $SL(2, Z)$ symmetry}

Let us first briefly mention how $SL(2, Z)$ symmetry appears 
in string theory. Evidence of $SL(2, Z)$ symmetry (S-duality) 
was first seen in the toroidal compactification of heterotic 
string to four dimensions. This is a symmetry of heterotic 
string equations of motion \cite{schsen,sen}. It is manifest 
in Einstein frame, and a subgroup of these transformations 
inverts the string coupling constant and simultaneously 
interchanges electric and magnetic fields. 

Type II string compactified on 
$K_3 \times T^2$ also exhibits this symmetry as a strong-weak 
coupling duality in four dimensions \cite{hultow}. 
The geometric realisation of this symmetry in both heterotic 
case as well as type II case was achieved through 
the string-string duality conjecture in six dimensions, 
between heterotic string on $T^4$ and type IIA string on 
$K_3$. On toroidal compactification, both of them have 
$SL(2, Z)$ as the S-duality symmetry which can now be 
interpreted as modular group of internal torus in 
the following manner. Torus compactification of both 
of these six dimensional theories to four dimensions 
gives $O(2, 2, Z)$ T-duality group. $O(2, 2, Z)$ 
is isomorphic to $SL(2, Z) \times SL(2, Z)$, which act on 
the complex structure modulus and the K\" ahler modulus of 
a torus respectively. It was shown in \cite{tri} that for 
a fixed K\" ahler modulus the S-duality group of heterotic 
string is the modular group ($SL(2, Z)$ corresponding to 
the complex structure deformation of torus) of the torus on 
which type IIA string is compactified. Similarly the S-duality 
group of type IIA string is modular group of the torus on 
which heterotic string  is compactified. Six dimensional 
string-string duality thus helps us establish the cross 
connection between S-duality of one string theory with 
the T-duality of another allowing us to give geometric 
interpretation of S-duality symmetry of four dimensional 
string theory.

Another string theory which has $SL(2, Z)$ S-duality symmetry 
is type IIB string theory in ten dimensions. This symmetry 
mixes the fields in the NS and RR sector and leaves 
the equations of motion invariant. Under $SL(2, Z)$ 
transformations, the Einstein frame metric (see below) 
and the self-dual four form field are invariant; and, 
\begin{equation}\label{omega}
\lambda \to \frac{a \lambda + b}{c \lambda + d} \; , \; \; \; 
\quad 
\pmatrix{C_{\alpha \beta} \cr B_{\alpha \beta}} 
\to (\Omega^T)^{- 1} 
\pmatrix{C_{\alpha \beta} \cr B_{\alpha \beta}} \; , 
\end{equation}
where $\lambda =\lambda_1 + i\lambda_2 \equiv \chi 
+ i e^{- \phi}$ and 
\[
\Omega = \pmatrix{a & b \cr 
                  c & d}     \in SL(2, Z) \; . 
\]
The above transformation of $\lambda$ can also be written 
as ${\cal M} \rightarrow \Omega {\cal M} \Omega^T$, where 
and ${\cal M}$ is defined in equation (\ref{curlym}). 
This $SL(2, Z)$ symmetry of the ten dimensional type IIB 
strings had no natural geometric interpretation as a modular 
group of internal torus. In nine dimensions, however, it can be related 
to two torus compactification of M-theory \cite{schwarz}. 

In the previous section we saw that the D-3-brane action can 
be interpreted as the Nambu-Goto action for a fundamental 
3-brane whose target space is twelve dimensional  
with (11, 1) signature. The internal two dimensional metric is 
covariant under $SL(2, Z)$ transformations. 
As mentioned earlier $SL(2, Z)$ symmetry of type IIB string is 
manifest in the Einstein frame. We will therefore first absorb 
the dilaton factor and write the ten dimensional metric in 
the Einstein frame. The ten dimensional Einstein frame metric 
is given by 
\begin{equation}
g_{\mu\nu}^E = e^{-\phi/2} g_{\mu\nu}.
\end{equation}
Substituting this in (\ref{strmet}) we see that 
the Nambu-Goto action is manifestly invariant under 
$SL(2, Z)$ transformations (\ref{omega}) if 
${\bf Z_\alpha} \to (\Omega^T)^{-1} {\bf Z_\alpha}$. 
In the target space, the metric on the internal torus 
$g_{int} = {\cal M}$ (see (\ref{gint})) transforms under 
$SL(2, Z)$ as 
\begin{equation}
g_{int} \rightarrow \Omega g_{int} \Omega^T \; , 
\; \; \; \; \Omega\in SL(2, Z)
\end{equation}
and the toroidal coordinates $U^1$ and $U^2$ as well as 
the two forms $B_{\mu \nu}$ and $C_{\mu \nu}$ transform as, 
\begin{equation}
\pmatrix{U^1 \cr U^2} \rightarrow (\Omega^T)^{-1} 
\pmatrix{U^1 \cr U^2}\; ,\;
\pmatrix{C_{\mu \nu} \cr B_{\mu \nu}} \rightarrow 
(\Omega^T)^{-1} \pmatrix{C_{\mu \nu} \cr B_{\mu \nu}}\; .
\end{equation}
We thus see that $SL(2, Z)$ symmetry of type IIB string theory 
in ten dimensions has a direct geometric interpretation in 
terms of the modular group of the internal torus. 
The $SL(2, Z)$ modulus of type IIB string theory in ten 
dimension is the complex structure modulus of this torus. 
Definition of torus coordinates $U^1$ and $U^2$, given in 
(\ref{u1u2}), involves $B$ and $C$ fields and,  
therefore, the twelve dimensional space can be interpreted as 
an elliptic fibration. If we set both $B$ and $C$ to zero, 
then $U^1 = V$ and $U^2 = U$. In 
this case the twelve dimensional space is a direct product of 
ten dimensional spacetime and the internal torus. The fact 
that this toroidal compactification does not involve second 
$SL(2, Z)$ alluded to earlier in this section implies that 
this torus has a fixed K\" ahler class. This also fits in 
nicely with the construction of 24 seven branes, and their 
relation to compactification of F-theory on elliptically 
fibered $K_3$ manifolds, given by Vafa \cite{v}. By showing 
the relation of $SL(2, Z)$ with modular group of the internal 
torus we have put this symmetry of type IIB strings on 
the same footing as the $SL(2, Z)$ S-duality symmetry of 
four dimensional $N=4$ supersymmetric string theory. 

We would like to note here an interesting implication of 
this interpretation of $SL(2, Z)$ symmetry. Strong-weak 
coupling duality transformation in type IIB string theory 
which is a subgroup of the $SL(2, Z)$ duality, exchanges 
the torus coordinates $U^1$ and $U^2$. As seen in section 3, 
the two radii of the torus are proportional to $e^{-\phi}$ 
and $e^{\phi}$, where $e^{- \phi}$ is the string coupling. 
We thus see that both in strong coupling or weak coupling 
limit, this theory appears to be eleven dimensional theory 
with $(10,1)$ signature. Perhaps, this is related to 
the conjectured self-duality of M-theory \cite{dvv}.

\section{Discussion}

To summarise, we start from the type IIB Dirichlet 3-brane 
action. Performing first a double dimensional reduction, and 
then applying the methods of \cite{s,t}, we obtain 
a Nambu-Goto action. It is interpreted as the world volume 
action of a fundamental 3-brane, and its target space theory 
as F-theory. We find that the target space is twelve 
dimensional, with signature $(11, 1)$. Locally, it is 
a product of a ten dimensional spacetime and a torus, with 
a fixed K\" ahler class. That is, the twelve dimensional space 
is an elliptic fibration over a ten 
dimensional base space \cite{v}. 

Also, the $SL(2, Z)$ symmetry of type IIB string has 
an explicit geometric interpretation: the $SL(2, Z)$ modulus 
of type IIB string is the complex structure modulus of 
the torus fiber. Moreover, the two radii of the torus are 
proportional to $e^{-\phi}$ and $e^{\phi}$, where 
$e^{- \phi}$ is the string coupling. Hence, both in the strong 
and weak coupling limit, the twelve dimensional theory appears 
to be eleven dimensional with (10,1) signature, which is 
perhaps related to the conjectured self-duality of M-theory. 

However, because of the double dimensional reduction, 
the Nambu-Goto action appears in a gauge and, hence, 
it does not have the full twelve dimensional general 
coordinate invariance. We also showed, by a simple counting 
argument, that this is likely to be a generic phenomenon. 
This may actually be a boon in disguise, as it may be related 
to the way F-theory implements supersymmetry, circumventing 
Nahm's theorem regarding realisations of supersymmetry in 
spacetime with dimensions $> 11$. 

We conclude by pointing out a couple of aspects related to 
F-theory, whose understanding is likely to provide deeper 
insights. One needs to understand thoroughly 
the gauge fixing we encountered in section 2. 
For this purpose, perhaps, one may start from 
an action more general than, or altogether different from, 
the Born-Infeld action for type II D-3-branes, and which has  
enough number of fields to provide more degress of freedom. 
One natural place to start may be the following: The duality 
symmetry of the Born-Infeld action is manifest only at 
the level of equations of motion \cite{g}. Writing this 
action in a manifestly duality invariant way, say, in analogy 
with the Maxwell's case treated in \cite{schsen}, will 
naturally introduce new fields. Such an action, if found, may 
lead to the Nambu-Goto action for a fundamental 3-brane, 
without any gauge conditions. 

Note that the target spacetimes of fundamental string, 
2-brane, and 3-brane are, respectively, 10, 11, and 12 
dimensional. In all these cases, the dual of the fundamental 
branes is always a 5-brane. The non trivial role of 3-branes 
and 5-branes in F-theory is also seen in \cite{sagn} from 
current algebra considerations. These observations suggest 
that perhaps 5-brane is the really fundamental object to 
study. In such an event, its study is likely to reveal 
important facts common to all these theories. A similar point 
of view is being advocated in \cite{dvv} also. 

\vskip 1cm

{\bf Note Added:} F-theory is expected to lead to M-theory 
upon double dimensional reduction. Hence, our double 
dimensional reduction 
procedure, adopted here for the sake of clarity and 
simplicity, may appear to imply that one is dealing with 
2-brane and M-theory, and not with 3-brane and F-theory. 

However, this is not the case. The two extra coordinates 
$U$ and $V$ that emerge in our approach are distinct 
from any of the original ten coordinates $X^\mu$, leading 
thus to the twelve dimensional F-theory. Also, only for 
the four dimensional 3-brane action (\ref{bi}) do the string 
coupling factors $e^\phi$ appear with the coefficients as 
given in (\ref{phis}) and (\ref{replace}). As seen in 
sections 3 and 4, it is these coefficients which eventually 
lead to the correct geometric interpretation of type IIB 
$SL(2, Z)$ symmetry. 

If one were dealing with 2-brane action then 
the $e^{- \frac{\phi}{2}}$ factor in equation (\ref{phis}) 
would be replaced by $e^{- \frac{2 \phi}{3}}$. The $e^\phi$ 
factors in (\ref{replace}) would also be replaced 
correspondingly. Then, the resultant Nambu-Goto action would 
not lead to a direct geometric interpretation of type IIB 
$SL(2, Z)$ symmetry. 

\vskip 1cm

{\bf Acknowledgement:} We thank A. Sagnotti and J. H. Schwarz 
for pointing out \cite{sagn} and \cite{sch} respectively. 
Also, we thank A. Sen and the referee for their critical 
comments about the emergence of F-theory. We hope that 
the Note Added clarifies this point better. D.J. would like 
to thank The Institute of Mathematical Sciences, Madras for 
kind hospitality where part of the work was done.

\end{document}